\documentclass[acmsmall]{acmart}

\usepackage{booktabs} 
\usepackage{graphicx}
\usepackage{yfonts}
\usepackage{url}
\usepackage{algorithm}
\usepackage{algorithmic}
\usepackage{subfigure}
\usepackage{comment}
\usepackage{booktabs}
\usepackage{multirow}
\usepackage{threeparttable}
\usepackage{supertabular}
\usepackage{diagbox}

\begin{document}

\title{Person-Job Fit: Adapting the Right Talent for the Right Job with Joint Representation Learning}

\author{Chen Zhu}
\affiliation{%
  \institution{Talent Intelligence Center, Baidu, Inc.}
  \city{Beijing}
  \country{China}}
\email{zhuchen02@baidu.com}
\author{Hengshu Zhu$^{*}$}
\affiliation{%
  \institution{Talent Intelligence Center, Baidu, Inc.}
  \city{Beijing}
  \country{China}}
\email{zhuhengshu@baidu.com}
\author{Hui Xiong$^{*}$}
\affiliation{%
  \institution{Business Intelligence Lab, Baidu Research}
  \city{Beijing}
  \country{China}}
\email{hxiong@rutgers.edu}
\author{Chao Ma}
\affiliation{%
  \institution{Talent Intelligence Center, Baidu, Inc.}
  \city{Beijing}
  \country{China}}
\email{machao13@baidu.com}
\author{Fang Xie}
\affiliation{%
  \institution{Talent Intelligence Center, Baidu, Inc.}
  \city{Beijing}
  \country{China}}
\email{xiefang@baidu.com}
\author{Pengliang Ding}
\affiliation{%
  \institution{Talent Intelligence Center, Baidu, Inc.}
  \city{Beijing}
  \country{China}}
\email{dingpengliang@baidu.com}
\author{Pan Li}
\affiliation{%
  \institution{Talent Intelligence Center, Baidu, Inc.}
  \city{Beijing}
  \country{China}}

\acmJournal{TMIS}



\begin{abstract}
Person-Job Fit is the process of matching the right talent for the right job by identifying talent competencies that are required for the job. While many qualitative efforts have been made in related fields, it still lacks of quantitative ways of measuring talent competencies as well as the job's talent requirements. To this end, in this paper, we propose a novel end-to-end data-driven model based on Convolutional Neural Network (CNN), namely Person-Job Fit Neural Network (PJFNN), for matching a talent qualification to the requirements of a job. To be specific, PJFNN is a bipartite neural network which can effectively learn the joint representation of Person-Job fitness from historical job applications. In particular, due to the design of a hierarchical representation structure, PJFNN can not only estimate whether a candidate fits a job, but also identify which specific requirement items in the job posting are satisfied by the candidate by measuring the distances between corresponding latent representations. Finally, the extensive experiments on a large-scale real-world dataset clearly validate the performance of PJFNN in terms of Person-Job Fit prediction. Also, we provide effective data visualization to show some job and talent benchmark insights obtained by PJFNN.
\end{abstract}

\maketitle

\let\thefootnote\relax\footnotetext{*Corresponding Author}

\section{Introduction}
Person-Job Fit refers to the process of matching the right talent for the right job through effectively linking talent competencies to job requirements. Many studies have shown that Person-Job Fit can be related to productivity and commitment~\cite{robbins2001organizational}. However, since there are huge numbers of job candidates and job postings available on the Internet, the gap between talent and job opportunities has been increasing regardless of the importance of Person-Job Fit.  For example, as of 2015, there were more than 400 million people available at LinkedIn~\cite{linkedinWiki}. Meanwhile, it is reported that recruiters need to averagely spend 52 days and 4,000 dollars for filling an open job position with a right talent~\cite{bersin}.

\begin{figure}
\vspace{0mm}
\includegraphics[width=\columnwidth]{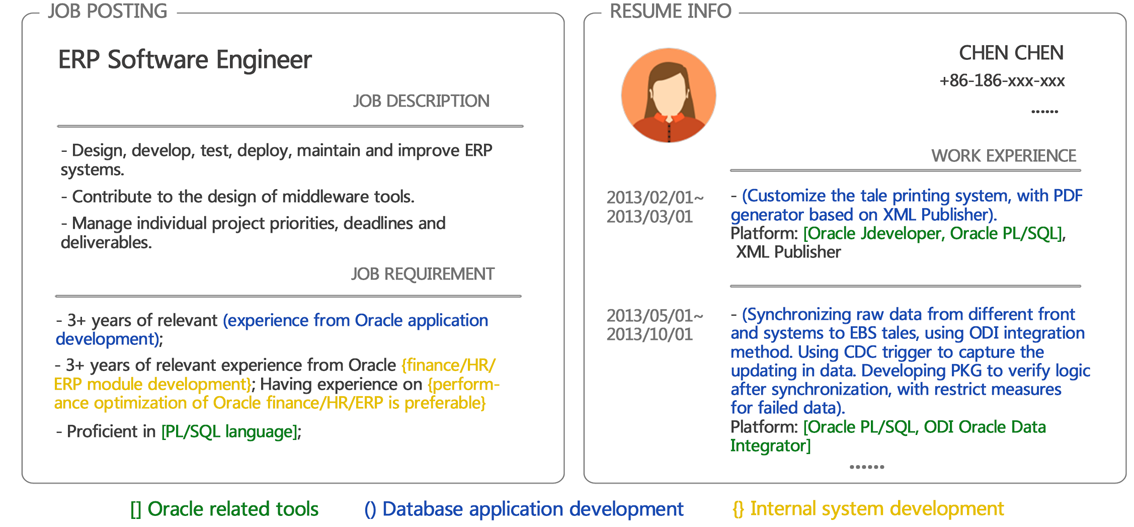}\\
\vspace{-2mm}
\caption{A motivating example of Person-Job Fit.}\label{jdResumeEp}
\vspace{-5mm}
\end{figure}
In the literature, the research related to Person-Job Fit is usually formulated as the problem of job/candidate recommendations~\cite{malinowski2006matching,zhang2016glmix,paparrizos2011machine,zhang2014research} and talent sourcing~\cite{zhu2016recruitment,xu2016talent}.
However, a significant challenge along this line is how to quantitatively measure talent competencies as well the job's talent requirements.
Figure~\ref{jdResumeEp} shows a motivating example of this paper. In the figure, a job posting and the resume of a successful applicant are listed. Specifically, there are three major requirements for this job which represent different needs of work responsibility, namely \emph{``Oracle related tools''}~(marked with ``[ ]''), \emph{``Internal system development''}~(marked with ``\{\}''), and \emph{``Database application development''}~(marked with ``( )''). For the applicant, the work experiences clearly validate that she has strong background in Oracle development and database applications, which match the job requirements well. As a result, while the applicant has no experience about internal system development, the employers believe she fits the job in general.
Indeed, while the above inspection of Person-Job fitness is apparent for those experts in this field, most recruiters, who are responsible for sourcing resumes, are not very familiar with the related knowledge and cannot justify the fitness efficiently.
If we can provide an explicable way of Person-Job Fit, it will help to improve the efficiency of recruiters.

To this end, in this paper, we propose a novel end-to-end data-driven model based on Convolutional Neural Network (CNN), namely Person-Job Fit Neural Network (PJFNN), to match the right talent for the right job by learning the joint representation of Person-Job fitness from historical job applications. Specifically, it assumes that both job postings and resumes can be projected onto a shared latent representation, along with each requirement item in job postings and each work experience item in resumes. Meanwhile, each job posting (requirement item) and its corresponding resumes (work experience items) should be similar in the representation space. With the help of PJFNN, we can not only estimate whether a candidate fits a job, but also identify which specific requirement items in the job posting are satisfied by the candidate. The major contributions of this paper can be summarized as follows.

\begin{itemize}
  \item We formulate Person-Job Fit as a joint representation learning problem, which aims to match a candidate's work experience to job requirement and thus provides a new research paradigm in talent recruitment.
\vspace{1mm}
  \item We propose a novel CNN-based end-to-end model, namely PJFNN, for the proposed problems, which projects both job postings and candidate resumes onto a shared latent representation by joint representation learning from historical job applications, along with each requirement item in jobs and each work experience item in resumes.
\vspace{1mm}
  \item The proposed method has been practiced in real-world scenarios. The extensive experiments on a large-scale real-world dataset clearly validate the performance of our approach in terms of Person-Job Fit prediction. Besides, we design effective data visualization to show some job and talent benchmark insights obtained by our method.
\end{itemize}

{\bf Overview.} The remainder of this paper is organized as follows. Section 2 provides a brief review of related works. Section 3 introduces the details of our model PJFNN. In Section 4, we report the evaluation results based on a real-world data set. Finally, we conclude the paper in Section 5.

\section{Related Work}
In this section, we will briefly introduce some works related to this paper.
According to the research problem and the technology used in this paper, the related works can be grouped into two categories, namely Person-Job Fit and text mining with neural network.

\subsection{Person-Job Fit}
Recruitment is a core function of human resource management. And the traditional effort to measure the fitness between employees and job positions is best articulated in Personality-Job Fit theory~\cite{robbins2001organizational}, which identifies six personality types~(i.e., Realistic, Investigative, Artistic, Social, Enterprising, and Conventional)~\cite{holland1973making} and proposes that the fitness between personality type and occupational environment determines the job satisfaction and turnover. Although this theory has been widely accepted in academia and industry world, how to precisely measure personality and fitness between jobs and jobs seekers is a vital problem every recruiter will face. Traditionally, a person's personality profile is measured by a well-designed inventory questionnaire, and the fitness is determined by recruiters without objective metric. Obviously, this method is subjective and often leads to biases.

Due to the explosion of online recruitment markets, recruitment analysis has been attracting more and more attentions~\cite{zhu2016recruitment,xu2018measuring,qin2018Enhancing} from researchers.
Traditional efforts tend to treat Person-Job Fit as a job/candidate recommendation problem.
In 2006, Malinowski~\emph{et al.} tried to find a good match between talents and jobs by two distinct recommendation systems~\cite{malinowski2006matching}.
Then Diaby~\emph{et al.} presented a content-based recommender system for recommending jobs to Facebook and LinkedIn users~\cite{diaby2013toward}.
Lu~\emph{et al.} exploited job and user profiles and the actions undertaken by users to propose a hybrid recommender system~\cite{lu2013recommender}.
To address the challenge that job applicants do not update the user profile in a timely manner, Wenxing~\emph{et al.} extended users' profile dynamically by job application records and their behaviors for better recommendation~\cite{hong2013dynamic}.
Zhang~\emph{et al.} leveraged collaborative filtering and some background information to recommended suitable jobs for candidates~\cite{zhang2014research}.

Recently, some researchers tried to study the Person-Job Fit problem from novel perspectives. For example, Paparrizos~\emph{et al.} exploited all historical job transitions as well as the data associated with employees and institutions to predict an employee's next job transition~\cite{paparrizos2011machine}. In~\cite{zhang2016glmix}, Zhang~\emph{et al.} created the generalized linear mixed model (GLMix), a more fine-grained model at the user or item level, for LinkedIn job recommender system and generated 20\% to 40\% more job applications. Li~\emph{et al.} proposed an approach to apply standardized entity data to improve job search quality in LinkedIn and to make search results more personalized~\cite{Li2016How}.
In~\cite{cheng2013jobminer}, job information is extracted from social network and used to construct an inter-company job-hopping network, which clearly demonstrates the flow of talents. Xu~\emph{et al.} measured the popularity of job skill by modeling the generation of skill network~\cite{xu2018measuring}. Lin~\emph{et al.} proposed to collaboratively model both textual (e.g., reviews) and numerical information to learn the latent structural patterns of companies~\cite{Lin2017Collaborative}. Shen~\emph{et al.} tried to improve recruitment efficiency by intelligent interview assessment~\cite{shen2018joint}.

Although the above studies have explored different research aspects of Person-Job Fit, few of them can provide comprehensible reasons behind their job/candidate recommendation results, which benefit both employers and job seekers.

\subsection{Text Mining with Neural Network}
Because both of resumes and job postings are textual data, Person-Job Fit can be treated as match between texts. Recently, deep neural network~(DNN) has become one of the hottest techniques in this field due to its good performance.

DNNs applied in text mining can be generally divided into two categories, convolutional neutral network~(CNN) and recurrent neural network~(RNN).
CNN aims at modelling hierarchical relationships and extracting local semantics. The effort of applying CNN in text mining can date back to~\cite{kalchbrenner2014convolutional}, where Kalchbrenner~\emph{et al.} creatively proposed Dynamic Convolutional Neural Network~(DCNN) to modelling sentences. Then many researchers began to solve NLP problems by CNN. In \cite{kim2014convolutional}, Kim {\it et al.} demonstrated that CNN, even just using a convolutional layer, also performs remarkably well in many NLP tasks.
Different from CNN, RNN is good at modelling sequence relationships and finding global semantics. Thus it performed very well in sequential labeling problems in text mining, such as machine translation~\cite{sutskever2014sequence} and contextual parsing~\cite{vinyals2015grammar}.

In text mining, both of machine translation and multilingual word embedding study on aligned data and thus are similar to our problem. Neural machine translation, which aims to build a neural network to read a sentence and output a correct translation, is a newly emerging approach to machine translation. Most of these works belong to a family of Encoder-Decoder. For example, Sutskever~\emph{et al.} proposed to use LSTMs to map sentences to sentences~\cite{sutskever2014sequence}. In~\cite{cho2014properties}, Cho~\emph{et al.} proposed a novel gated recursive convolutional neural network for Encoder-Decoder based translation framework. Devlin~\emph{et al.} proposed a novel formulation for neural network joint model and yielded strong empirical results~\cite{devlin2014fast}.
On the other hand, Multilingual word embedding aims to map words from different languages into a shared latent space. In~\cite{Lauly2014Learning}, Lauly~\emph{et al.} tried to use an autoencoder to learn word representations. Similarly, Hermann~\emph{et al.} tried to assign similar embeddings to aligned sentences for learning semantic representations without aligned words~\cite{hermann2014multilingual2}.

Obviously, both neural machine translation and multilingual word embedding are good solutions for learning the relationships between aligned data, which also exist in Person-Job Fit problem. However, most of these methods, such as~\cite{hermann2014multilingual2}, need aligned relationships in sentence level or even word level, which is not available in our problem.
In this paper, we modify some state-of-the-art ideas in the above works to adapt our problem and propose a novel model to link talents to jobs.

\section{Model Description}
In this section, we will first introduce some research preliminaries, and then explain the proposed model, PJFNN, in detail.

\subsection{Preliminaries}

In practice, the recruitment data usually consist of three parts, namely job postings, resumes and job application records. Specifically, a job posting contains \emph{job content} (e.g., job duty), and \emph{job requirement} that consists of several requirement items (e.g., qualifications about skills or experiences). A resume contains a candidate's \emph{profile} (e.g., age and gender), and work experience that consists of several work experience items (e.g., project experiences in previous/current companies).
To simplify our problem, we assume that a job can be represented by its job requirements, and the work experiences of a candidate can mainly reflect her competency.
Thus we can formulate Person-Job Fit as matching a candidate's work experiences to job requirements.
In particular, we regard a job posting as a set of requirement items, and a resume as a set of work experience items. Those job application records naturally provide labeled data for us. Please note that there exist overlaps in job application data. In other words, a candidates can apply for several jobs, and of course a job can be applied by many candidates. Thus in different job applications, the jobs may be the same. So may those resumes.

We assume that there is a shared latent representation for job postings and resumes. Accordingly, each job posting and each resume can be represented by a vector on the latent representation. Therefore, a candidate fits a job well only if their vectors on the representation are similar. Similarly, requirement items and work experience items can also be projected onto this representation, and the distances between them reflect the corresponding fitness. To facilitate understanding, we still take the case in Figure~\ref{jdResumeEp} as an example. Specifically, \emph{``Oracle related tools''}, \emph{``Internal system development''}, and \emph{``Database application development''} are three latent factors that jointly form the representation of the given job posting. We can observe that the two work experiences in resume endorse that the candidate fits the requirement 1 and 3 of this job posting well, but misses the requirement 2 related to \emph{``Internal system development''}. Therefore, compared with requirement 2, the latent vectors of both the work experience items should be more similar to those of the other requirements. Meanwhile, considering this candidate has met 2 of 3 requirements, the latent vectors of the job posting and the resume should be relatively similar.

Formally, in this paper, the successful job application records are represented by set $A$, and each record $a_i \in A$ is a pair $(j_i, r_i)$, where $j_i$ and $r_i$ are the corresponding job posting and resume respectively. Job posting $j_i$ is a set of items $j_i=\{j_{i,0}, j_{i,1}, ..., {j_{i,n_{j_i}}}\}$, where $n_{j_i}$ is the number of requirement items in $j_i$. Resume $r_i$ is also a set of work experience items $r_i=\{e_{i,0}, e_{i,1}, ..., e_{i,n_{r_i}}\}$. Besides, we use $F$ to represent the set of failed job applications and each record $f_i \in F$ is a pair $(j_f, r_f)$. We define $\bf{v^{j_i}}$ as the latent vector of job posting $j_i$ and $\{\bf{v^{j_i}_n}\}_n$ as the latent vectors of requirement items in this job posting. Furthermore, the latent vectors $\bf{v^{r_i}}$ and $\{\bf{v^{r_i}_n}\}_n$ are defined for resumes by a similar way.

\vspace{2mm}
\subsection{Person-Job Fit Neural Network}
\vspace{2mm}
\begin{figure*}[!t]
\includegraphics[width=\textwidth]{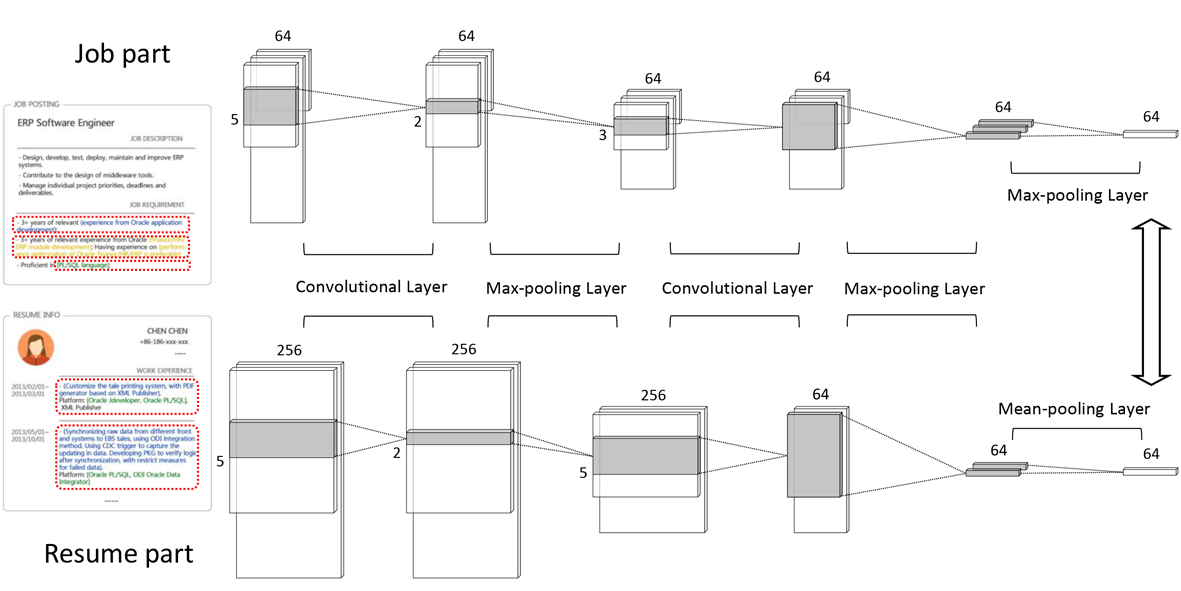}\\
\caption{An illustration of the architecture of PFJNN, which can be separated into two parts~(e.g., job part and resume part). Two one-dimensional convolutional layers are applied on each requirement~(work experience) item. Each convolutional layer is followed by a max-pooling layer, where the stride of the first max-pooling layers is 2, and the size of the second max-pooling layer is set as the length of its input to map a requirement~(work experience) item into a vector. The corresponding hyper-parameters are shown in this figure. Finally, the vectors of all of requirement~(work experience) items are projected onto a vector by a max-pooling layer~(mean-pooling layer) to represent corresponding job posting~(resume).}\label{PFJNN}
\end{figure*}

To match the right talent for the right job, we propose a CNN-based model, PJFNN, for effectively learning the joint representation of Person-Job fitness. Its architecture and hyper-parameters are shown in Figure~\ref{PFJNN}. Generally, PJFNN is a bipartite neural network that can be separated into two parallel parts, namely \emph{job part} and \emph{resume part}. With this design, PJFNN can project job postings and candidate resumes onto a shared latent representation respectively.
In the following, we will explain the job part in detail and then briefly introduce the resume part by pointing out the differences between them.

\textbf{Job Part:} As mentioned above, each job is regarded as a set of requirement items. Specifically, given an item with $|S|$ words, we take the embedding $\bf{w_{i}}\in R^{d}$ for each word and construct the corresponding matrix $S\in R^{d\times |S|}$,
\begin{equation}
S=[\bf{w_1},\bf{w_2},...,\bf{w_s}].
\end{equation}
The item matrix is the input layer of PJFNN, and a job posting $j_i$ can be represented as
\begin{equation}
j_i=[S_1,S_2,...,S_{n_{j_i}}]^{\top},
\end{equation}
where $n_{j_i}$ is the number of requirement items in job posting $j_i$.

Then we apply two one-dimensional convolutional layers on the input layer. The one-dimensional convolution is an operation between a vector of weights $\bf{m}$ and a vector of input viewed as a sequence $S$. The idea behind this operation is to take the dot product of the weight parameter $\bf{m}$ with each m-gram in a sequence to produce output sequence $\bf{C}$, i.e.,
\begin{equation}
c_{i}=\bf{m}^{\top}S_{i-|m|+1:i}.
\end{equation}
Because one-dimensional convolution can deal with unfixed-length sequence, it is widely used to model text~\cite{kalchbrenner2014convolutional}.

To reduce the training cost, we apply Batch Normalization~\cite{ioffe2015batch} on the outputs of one-dimensional convolutional layers. Batch Normalization is a mechanism for dramatically accelerating the training of deep networks, which also eliminates the importance of initialization. Furthermore, the Batch Normalization is followed by a Rectified Linear Unit~(ReLU) layer~\cite{nair2010rectified} and a one-dimensional max-pooling layer. Please note that the size of the second max-pooling layer is the length of its input, so that a sentence can be mapped into a vector, $\bf{v^{j_i}_{n}}$. With this part of PJFNN, job posting $j_i$ can be transformed into
\begin{equation}
[\bf{v^{j_i}_0}, \bf{v^{j_i}_1}, ..., \bf{v^{j_i}_{n_{j_i}}}]^{\top}.
\end{equation}
Then we use a max-pooling layer to integrate them into $\bf{v^{j_i}}$ as
\begin{equation}
\bf{v^{j_i}}=[max(v^{j_i}_{*,0}), max(v^{j_i}_{*,1}), ..., max(v^{j_i}_{*,l})],
\end{equation}
where $l$ is the length of latent requirement vectors, and $v^{j_i}_{*,k}$ is the vector of all representations of items in dimension $k$.

\textbf{Resume Part:} Generally, the resume part of PJFNN is similar to the job part. The difference is how to integrate item representations into resume representations. Here, we propose to use a mean-pooling layer for modelling them and set
\begin{equation}
\bf{v^{r_i}}=(v^{r_i}_0+v^{r_i}_1+...+v^{r_i}_{n_{r_i}})/n.
\end{equation}

To get a reasonable shared low-dimension representation, we minimize
\begin{equation}
Loss(A) = \sum^{|A|}_{i=1}D(j_i, r_i),
\end{equation}
where $D(j_i, r_i)$ is the distance between $v^{j_i}$ and $v^{r_i}$. However, if we directly minimize the loss function above, PJFNN would learn to reduce all parameters to zero. To address this issue, we simultaneously minimize the distances between representations of resumes $v^{r_i}$ and job postings $v^{j_i}$ in the successful job applications and maximize the distances between representations of resumes $v^{r_f}$ and job postings $v^{j_f}$ in the failed job applications. It is obvious that how to select the failed application set has large impact on the performance of our model. We will discuss this in detail in Section~\ref{sec:exp}. Besides, we also add $L_2$ regularization of all parameters in PJFNN $\theta$ to avoid over-fitting. Therefore, the objective function is formulated as
\begin{equation}
min(\sum^{|A|}_{i=1}D(j_i, r_i)-\sum^{|F|}_{f=1}D(j_f, r_f)+\lambda\|\theta\|^2).
\end{equation}
In this paper, we select cosine similarity to measure the distance between latent representations,
\begin{equation}
D(j_i, r_i) = -\frac{\bf{v^{j_i}} \cdot \bf{v^{r_i}}}{\|\bf{v^{j_i}}\|\|\bf{v^{r_i}}\|}.
\end{equation}
Finally, the objective function can be rewritten as
\begin{equation}
min(\sum^{|A|}_{i=1}-\frac{\bf{v^{j_i}} \cdot \bf{v^{r_i}}}{\|\bf{v^{j_i}}\|\|\bf{v^{r_i}}\|} + \sum^{|F|}_{f=1}\frac{\bf{v^{j_f}} \cdot \bf{v^{r_f}}}{\|\bf{v^{j_f}}\|\|\bf{v^{r_f}}\|}+\lambda\|\theta\|^2).
\end{equation}
The objective function is optimized by the Adam algorithm~\cite{Kingma2014Adam}.

\textbf{Discussion (CNN v.s. RNN).} In text mining, most of DNNs can be classified into two categories, CNN and RNN. There have been a lot of work on comparing their performance on text mining tasks~\cite{vu2016combining,yin2017comparative}. Generally, it is believed that CNN is good at modelling hierarchical relationships and local semantics. Meanwhile, RNN focuses on sequential dependency and global semantics. In PJFNN, we propose to use CNN rather than RNN for modelling textual data, which is because that CNN can better capture the hierarchical relationships and local semantic between a job posting (resume) and its requirement~(work experience) items.
Actually, in the recruitment data, each item only consists of short sentences and limited keywords, and the sequential dependency between different items are not significant. Therefore, we think CNN is a more effective approach for modelling the textual data in Person-Job Fit.

\textbf{Discussion (Job v.s. Resume).} In PJFNN, we propose to leverage the max-pooling for job part modelling and the mean-pooling for resume part modelling. Indeed, we think each dimension of the shared latent representation can reflect a specific latent aspect of expertise in some ways. Intuitively, since the job posting is usually well-formatted, different requirement items in a job posting usually represent different aspects of expertise independently. In contrast, each work experience item of candidates is often a mixture of expertise, since candidates usually want to thoroughly demonstrate their abilities in the work experience. Just as the example shown in Figure~\ref{jdResumeEp}, three requirement items represent three different needs, containing \emph{``Oracle related tools''}, \emph{``Internal system development''}, and \emph{``Database application development''}, while both of the candidate's work experience items meet the first and the third needs of this job. 
\section{Experiments}\label{sec:exp}
In this section, we will evaluate the performance of our model on a large-scale real-world recruitment dataset.

\begin{figure*}[!t]
\centering
\subfigure[] {\label{fig:NofAppltoTime}
\includegraphics[width=0.23\textwidth]{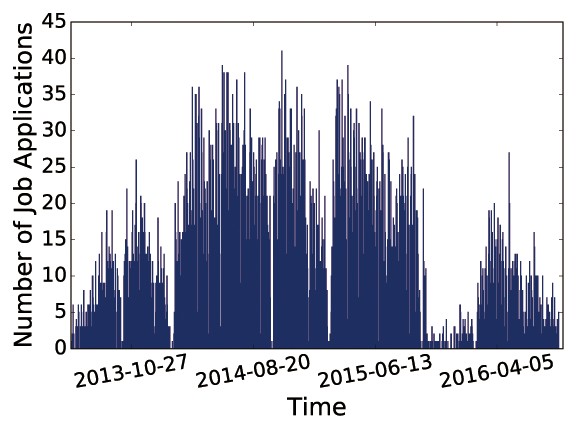}
}
\hspace{-2.5mm}
\subfigure[] {\label{fig:NofJobtoNofAppl}
\includegraphics[width=0.23\textwidth]{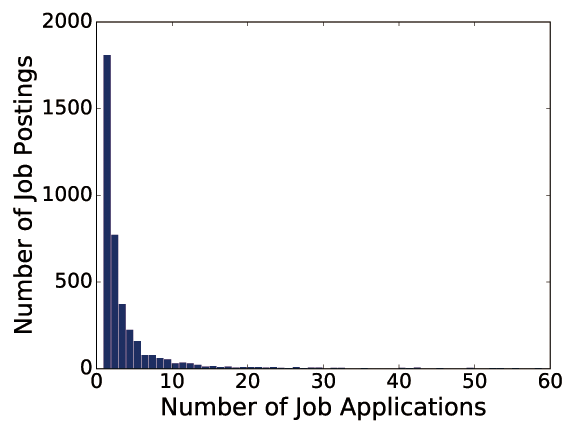}
}
\hspace{-2.5mm}
\subfigure[] {\label{fig:NofJobtoClass}
\includegraphics[width=0.23\textwidth]{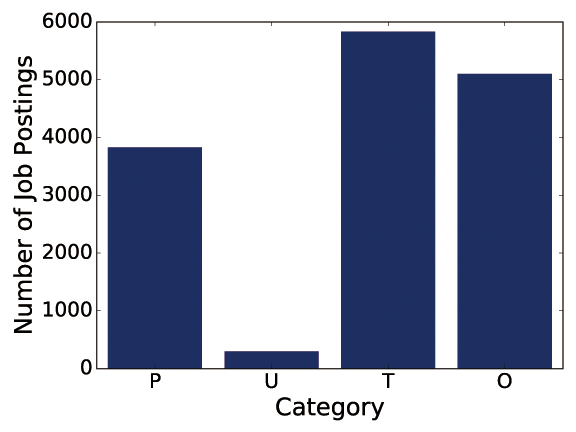}
}
\hspace{-2.5mm}
\subfigure[] {\label{fig:NofResumetoClass}
\includegraphics[width=0.23\textwidth]{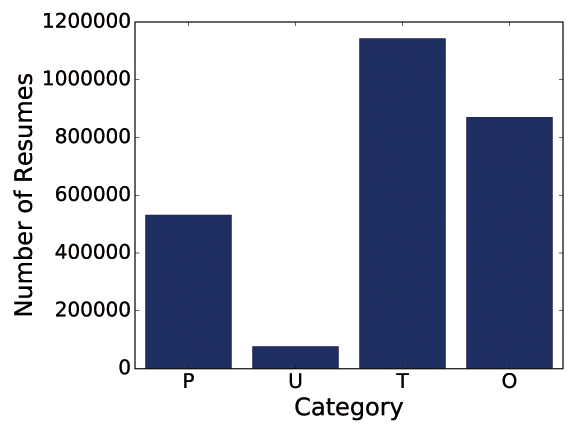}
}
\vspace{-3mm}
\caption{The distribution of (a) the number of successful job applications with respect to time spans, (b) the number of job postings with respect to the number of their successful job applications, (c) the number of job postings with respect to different categories, (d) the number of resumes with respect to different categories.}\label{dataDecs}
\vspace{-2mm}
\end{figure*}

\subsection{Experimental Setup}
The dataset used in the experiments is the historical job application records of a large high tech company in China, which ranges from 2013 to 2016.
It contains more than two million resumes and 15,039 job postings, while there are only 31,928 successful job applications. Indeed, the low admit rate ($\approx$ 1\%) clearly validates the importance of Person-Job Fit in talent recruitment. To avoid biases, we removed intern applications and resumes/job postings without any information. After that, the filtered dataset contains 12,007 successful job applications. Specifically, we demonstrate some basic statistics of our dataset in Figure~\ref{dataDecs}. From Figure~\ref{fig:NofAppltoTime}, we can observe that the number of recruitment is relatively steady, except from October 2015 to February 2016, where the dramatic decline is due to the change of recruitment policy (i.e., partial hiring freeze). Meanwhile, the number of job postings with respect to the number of their successful job applications roughly follows a long tail distribution according to Figure~\ref{fig:NofJobtoNofAppl}. In our dataset, each job posting/resume is classified into a job category by work content, which can be \emph{Technology~(T)}, \emph{Product~(P)}, \emph{User interface/experience~(U)}, or \emph{Others~(O)}. From Fiqure~\ref{fig:NofJobtoClass} and Figure~\ref{fig:NofResumetoClass}, we can find that the recruitment demand of category T is the largest, followed by P and O.

What should be mentioned is that words should be represented by vectors firstly in our model. Thus we first used Skip-gram Model~\cite{Mikolov2013Efficient} to encode words of resumes and job postings into 64-dimension and 256-dimension vectors, respectively. Please note that, to get a generalized embedding, we trained the Skip-gram on the entire dataset rather than the filtered one. And the word embedding would not be changed during training of our model.

Besides, all of applicants in successful applications must go through rigorous interview, thus they can be directly used as positive samples. But as for these failed job applications, the reasons of these failures are various. For example, some failed applications may be just due to the low pay/benefits rather than the unfitness of their work experiences and job requirements. These applications, which were labeled as failure, should be regarded as successful applications.
Thus, to guarantee the effectiveness of representation learning, we generated negative samples by replacing job postings in successful applications with randomly selected job postings, instead of extracting negative samples from failed applications directly. Moreover, to evaluate the performance of PJFNN fairly, we conducted experiments on both a semi-synthetic data and the real-world data.

\subsection{Evaluation on Person-Job Fit Prediction}
Here, we will evaluate PJFNN by predicting whether an applicant fits a job, namely, the fitness between a given job posting and a resume.

Specifically, because PJFNN, which is for representation learning, does not apply to prediction problems directly,
here we use the cosine similarities between representations of job postings and resumes learned by PJFNN as its prediction results.

To construct the dataset, we generate the same number of negative samples as that of positive samples.
Along this line, we randomly selected 80\% of the dataset as training data, another 10\% for tuning the parameters, and the last 10\% as test set to validate the performance.
To prove the effectiveness of PJFNN and our ideas about failed applications, we will demonstrate the performances of PJFNN and baselines, based on a semi-synthetic data (i.e., negative samples are manually generated) and the real-world data (i.e., negative samples are randomly sampled from the failed application records). Besides, we also splited our dataset according to time and validate the performance of our model in each year.

\textbf{Baselines Methods.} We first selected several classic classification methods as baselines, including \emph{Logistic Regression~(LR)}, \emph{Decision Tree~(DT)}, \emph{Naive Bayes~(NB)}, \emph{Support Vector Machine~(SVM)}, \emph{Adaboost~(Ada)}, \emph{Random Forests~(RF)}, \emph{Gradient Boosting Decison Tree~(GBDT)}, \emph{Linear Discriminant Analysis~(LDA)}, and \emph{Quadratic Discriminant Analysis~(QDA)}. For these baselines, it is unreasonable to directly use word vectors as input, which will lead to the curse of dimensionality. Thus, we treated the mean vector of all word vectors in a resume~(job posting) as its latent vector, and then regard the latent vectors of a candidate's resume and the corresponding job posting together as the input of baseline methods. It is obvious that the quality of word representation has large impact on the prediction performance. However, since both of our model and these baselines are based on the same word representation, we think its quality cannot affect the validation of our model's effectiveness. Thus we did not conduct experiments with different word embedding models and different dimensions of word representations.

Besides, because PJFNN is to map resume~(job posting) onto a shared representation, it can also be treated as a dimensionality reduction method. Thus we selected a dimensionality reduction method, \emph{PLTM}~\cite{mimno2009polylingual}, as a baseline, which can discover topics aligned across multiple language. Here, we treat resumes and job postings as different languages and represent both of them as distributions over a shared set of topics learned by PLTM. We calculated the cosine similarities between the distributions of job postings and resumes as its prediction results.

\textbf{Evaluation Metrics.} We evaluate the model performance by area under curve of ROC, or \emph{AUC} for short, instead of other classic metrics for classification, such as \emph{precision}, \emph{recall}, and \emph{F1}. It is because that AUC can reflect models' performance within different boundary value between classes and thus is widely used in two-class classification problems.

\subsubsection{Evaluation on Semi-synthetic Data}
Here the positive samples are real successful applications in our dataset and the negative samples are manually generated by randomly selecting resumes and job postings.
We demonstrate the AUC and significance test results in Table~\ref{tab:AUC}. Obviously, PJFNN has the best performance. Besides, PLTM, random forests and GBDT also perform relatively well.

Besides, we splited the dataset by time and train models separately on them for further evaluating our model within each year. The results are shown in Table~\ref{tab:AUC_forYear}. We can find that our model also outperforms all baselines in all of years, which proves the effectiveness of PJFNN, again. Besides, the performances of all methods in 2014 and 2015 are obviously better than those in 2013 and 2016. We think it may be due to the lack of data. Actually, the numbers of successful applications in 2013-2016, which are 1403, 5492, 3903, and 1209, are highly correlated with the corresponding AUC results.

The corresponding AUC results for different job categories are listed in Table~\ref{tab:AUC_forTPUO}. Therein the performances of all methods are relatively poor in \emph{O}. We think it may be due to candidates' resumes cannot well reflect the skills required by jobs of \emph{O}~(e.g. marketing and customer service). After all, compared with other categories, where jobs are highly related to some professional skills~(e.g. coding, program management, and UI design), jobs in \emph{O} often do not have clear skill requirements and thus it is hard to find suitable candidates for them just by resumes.

\begin{table}[]
\vspace{-0mm}
 \caption{The AUC performance of PFJNN and baselines on semi-synthetic data.}
 \scriptsize
 \centering
 \begin{tabular}{c | c  c  c  c  c  c  c  c  c  c  c  c }
  \toprule
   & \textbf{PJFNN} & DT & LR & NB & SVM & LDA & QDA & Ada & RF & GBDT & PLTM \\
  \midrule
   AUC & \textbf{0.85026} & 0.65447 & 0.74757 & 0.67327 & 0.75241 & 0.76017 & 0.56764 & 0.71552 & 0.80104 & 0.83140 & 0.81568 \\
   Improve & -- & +29.91\% & +13.37\% & +26.28\% & +13.00\% & +11.85\% & +49.78\% & +18.83\% & +6.14\% & +2.26\% & +4.23\% \\
   P-Value & -- & <0.001 & <0.001 & <0.001 & <0.001 & <0.001 & <0.001 & <0.001 & <0.001 & 0.001 & <0.001 \\
  \bottomrule
 \end{tabular}\label{tab:AUC}
\end{table}

\begin{table}[]
\vspace{-0mm}
 \caption{The AUC performance of PFJNN and baselines on semi-synthetic data in each year.}
 \tiny
 \centering
 \begin{tabular}{c | c  c  c | c  c  c | c  c  c | c  c  c }
  \toprule
  \multirow{2}{*}{Method} & \multicolumn{3}{c|}{2013} & \multicolumn{3}{|c}{2014}  & \multicolumn{3}{|c}{2015}  & \multicolumn{3}{|c}{2016} \\
                          & AUC & Improve & P-Value & AUC & Improve & P-Value & AUC & Improve & P-Value & AUC & Improve & P-Value \\
  \midrule
   \textbf{PJFNN} & \textbf{0.81891} & -- & -- & \textbf{0.86272} & -- & -- & \textbf{0.84486} & -- & -- & \textbf{0.81990} & -- & -- \\
   DT & 0.58573 & +39.81\% & <0.001 & 0.68980 & +25.06\% & <0.001 & 0.63743 & +32.54\% & <0.001 & 0.55047 & +48.94\% & <0.001 \\
   LR & 0.73316 & +11.69\% & <0.001 & 0.75874 & +13.70\% & <0.001 & 0.72085 & +17.20\% & <0.001 & 0.74339 & +10.29\% & <0.001 \\
   NB & 0.66495 & +23.15\% & <0.001 & 0.67823 & +27.20\% & <0.001 & 0.63483 & +33.08\% & <0.001 & 0.61375 & +33.58\% & <0.001 \\
   SVM & 0.72413 & +13.08\% & <0.001 & 0.75348 & +14.49\% & <0.001 & 0.73071 & +15.62\% & <0.001 & 0.69660 & +17.70\% & <0.001 \\
   LDA & 0.71475 & +14.57\% & <0.001 & 0.75627 & +14.07\% & <0.001 & 0.72081 & +17.20\% & <0.001 & 0.70405 & +16.45\% & <0.001 \\
   QDA & 0.70876 & +15.54\% & <0.001 & 0.65188 & +32.34\% & <0.001 & 0.52678 & +60.38\% & <0.001 & 0.49262 & +66.43\% & <0.001 \\
   Ada & 0.67320 & +21.64\% & <0.001 & 0.72488 & +19.01\% & <0.001 & 0.70941 & +19.09\% & <0.001 & 0.63677 & +28.75\% & <0.001 \\
   RF & 0.69840 & +17.25\% & <0.001 & 0.80338 & +7.38\% & <0.001 & 0.79950 & +5.67\% & <0.001 & 0.73915 & +10.92\% & <0.001 \\
   GBDT & 0.76122 & +7.57\% & <0.001 & 0.83079 & +3.84\% & 0.004 & 0.82405 & +2.52\% & 0.038 & 0.77446 & +5.86\% & <0.001 \\
   PLTM & 0.79520 & +2.98\% & <0.001 & 0.81091 & +6.38\% & <0.001 & 0.80145 & +5.41\% & <0.001 & 0.71000 & +15.47\% & <0.001 \\
  \bottomrule
 \end{tabular}\label{tab:AUC_forYear}
\end{table}

\begin{table}[]
\begin{minipage}{0.45\linewidth}
\vspace{-0mm}
 \caption{The AUC performance of PFJNN and baselines on semi-synthetic data in terms of job categories and years.}
 \tiny
 \centering
 \begin{tabular}{c|c|c|c|c}
  \toprule
  \diagbox[width=20mm,height=7mm]{Year}{Job category} & T & P & U & O \\
  \midrule
  2013 & 0.88267 & 0.87787 & 1.0 & 0.67677 \\
  2014 & 0.91797 & 0.88105 & 0.88194 & 0.75809 \\
  2015 & 0.90484 & 0.88596 & 0.73964 & 0.73606 \\
  2016 & 0.80120 & 0.90795 & 1.0 & 0.74122 \\
  overall & 0.89811 & 0.89165 & 0.84131 & 0.76599 \\
  \bottomrule
 \end{tabular}\label{tab:AUC_forTPUO}
\end{minipage}
\hspace{4mm}
\begin{minipage}{0.45\linewidth}
\vspace{-0mm}
 \caption{The AUC performance of PFJNN and baselines on real-world data  in terms of job categories and years.}
 \tiny
 \centering
 \begin{tabular}{c|c|c|c|c}
  \toprule
  \diagbox[width=20mm,height=7mm]{Year}{Job category} & T & P & U & O  \\
  \midrule
  2013 & 0.79548 & 0.84303 & 0.85000 & 0.73190 \\
  2014 & 0.85283 & 0.79980 & 0.87878 & 0.68566 \\
  2015 & 0.75338 & 0.73597 & 0.83333 & 0.65537 \\
  2016 & 0.71839 & 0.67382 & 0.85454 & 0.60533 \\
  overall & 0.77533 & 0.77313 & 0.74637 & 0.75367 \\
  \bottomrule
 \end{tabular}\label{tab:AUCRealData_forTPUO}
\end{minipage}
\end{table}

\subsubsection{Evaluation on Real-world Data}

Here both positive samples and negative samples are real records in our dataset. We kept the proportion of positive samples to negative samples by randomly selecting the same number of failed applications as negative samples. Similarly, the performance on the entire dataset and on data of each year are shown in Table~\ref{tab:AUCRealData} and Table~\ref{tab:AUCRealData_forYear}, respectively. Table~\ref{tab:AUCRealData_forTPUO} records the corresponding AUC results fir different job categories on real-world data. Although the AUC results are much less than those on the semi-synthetic data, PJFNN still generally outperforms these baselines. But what should be noted is that the performances of all methods are limited in data of 2015 and 2016. We think it can be attributed to the change of recruitment policy (e.g. partial hiring freeze) beginning from October 2015, which results that many excellent candidates cannot be enrolled although their qualifications fit the jobs well. Indeed, this situation further validates the reasonability of manually generating negative samples in some ways.

\begin{table}[]
\vspace{-0mm}
 \caption{The AUC performance of PFJNN and baselines on real-world data.}
 \scriptsize
 \centering
 \begin{tabular}{c | c  c  c  c  c  c  c  c  c  c  c  c }
  \toprule
   & \textbf{PJFNN} & DT & LR & NB & SVM & LDA & QDA & Ada & RF & GBDT & PLTM \\
  \midrule
   AUC & \textbf{0.75852} & 0.59903 & 0.70009 & 0.60660 & 0.69541 & 0.69237 & 0.48609 & 0.67598 & 0.67926 & 0.71507 & 0.57500 \\
   Improve & -- & +26.62\% & +8.34\% & +25.04\% & +9.07\% & +9.55\% & +56.04\% & +12.21\% & +11.66\% & +6.07\% & +31.91\% \\
   P-Value & -- & <0.001 & <0.001 & <0.001 & <0.001 & <0.001 & <0.001 & <0.001 & <0.001 & <0.001 & <0.001 \\
  \bottomrule
 \end{tabular}\label{tab:AUCRealData}
\end{table}

\begin{table}[]
\vspace{-0mm}
 \caption{The AUC performance of PFJNN and baselines on real-world data in each year.}
 \tiny
 \centering
 \begin{tabular}{c | c  c  c | c  c  c | c  c  c | c  c  c }
  \toprule
  \multirow{2}{*}{Method} & \multicolumn{3}{c|}{2013} & \multicolumn{3}{|c}{2014}  & \multicolumn{3}{|c}{2015}  & \multicolumn{3}{|c}{2016} \\
                          & AUC & Improve & P-Value & AUC & Improve & P-Value & AUC & Improve & P-Value & AUC & Improve & P-Value \\
  \midrule
   \textbf{PJFNN} & \textbf{0.77801} & -- & -- & \textbf{0.79549} & -- & -- & 0.67099 & -- & -- & 0.67486 & -- & -- \\
   DT & 0.65665 & +18.48\% & <0.001 & 0.58957 & +34.92\% & <0.001 & 0.59078 & +13.57\% & <0.001 & 0.62421 & +8.11\% & 0.186 \\
   LR & 0.73247 & +6.21\% & <0.001 & 0.67419 & +17.99\% & <0.001 & 0.66296 & +1.21\% & 0.102 & 0.58629 & +15.10\% & <0.001 \\
   NB & 0.69503 & +11.93\% & <0.001 & 0.63219 & +25.83\% & <0.001 & 0.56209 & +19.37\% & <0.001 & 0.58679 & +15.00\% & <0.001 \\
   SVM & 0.72833 & +6.82\% & <0.001 & 0.71769 & +10.84\% & <0.001 & 0.65552 & +2.35\% & 0.100 & 0.66165 & +1.99\% & 0.377 \\
   LDA & 0.65084 & +19.53\% & <0.001 & 0.68843 & +15.55\% & <0.001 & \textbf{0.67860} & -1.1\% & 0.697 & 0.65315 & +3.32\% & 0.981 \\
   QDA & 0.60720 & +28.13\% & <0.001 & 0.53379 & +49.02\% & <0.001 & 0.44534 & +50.66\% & <0.001 & 0.48284 & +39.76\% & <0.001 \\
   Ada & 0.59654 & +30.42\% & <0.001 & 0.65753 & +20.98\% & <0.001 & 0.63518 & +5.63\% & <0.001 & 0.64003 & +5.44\% & <0.001 \\
   RF & 0.69726 & +11.58\% & <0.001 & 0.65889 & +20.73\% & <0.001 & 0.66118 & +1.48\% & 0.127 & 0.65315 & +3.32\% & 0.06 \\
   GBDT & 0.69990 & +11.16\% & <0.001 & 0.71538 & +11.19\% & <0.001 & 0.65660 & +2.19\% & 0.576 & \textbf{0.67543} & -0.08\% & 0.581 \\
   PLTM & 0.56283 & +38.23\% & <0.001 & 0.54371 & +46.30\% & <0.001 & 0.56283 & +19.21\% & <0.001 & 0.54371 & +24.12\% & <0.001 \\
  \bottomrule
 \end{tabular}\label{tab:AUCRealData_forYear}
\end{table}

\subsection{Evaluation on Joint Representation Learning}
\begin{figure*}[t]
\centering
\includegraphics[width=\textwidth]{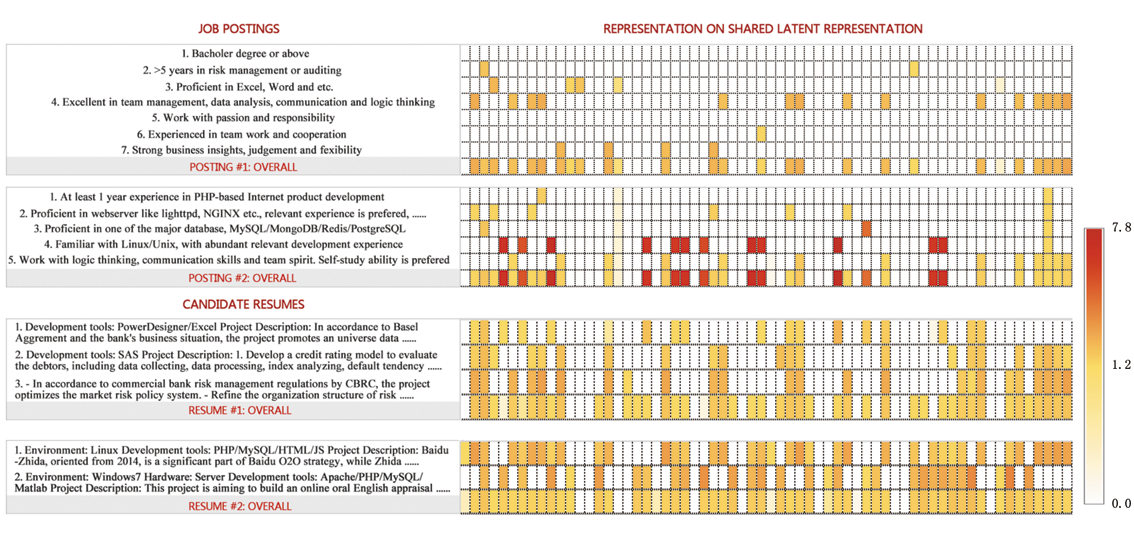}\\
\vspace{-3mm}
\caption{The representations of some resumes and job postings learned by PJFNN, where the darker color means the higher value. The representation of each requirement item~(work experience) and the overall representations of job postings~(resumes) are all shown.}\label{vectorCase}
\end{figure*}

Since the direct goal of PJFNN is to learn a shared latent representation for both resumes and job postings, here we show the representation vectors of some resumes and job postings for validating the effectiveness of joint representation learning. Specifically, we randomly selected two resumes and two job postings from different job categories, and their vectors learned by PJFNN are shown in Figure~\ref{vectorCase}, where the darker color means the higher value.

\textbf{Job Posting \#1.} Considering the job posting contains words ``risk management'', ``data analysis'', and ``business insights'', this job is clearly for recruiting people with risk management skills. However, compared with the requirement 2, which directly asks for ``$>$5 years in risk management or auditing'', the requirements 3, 4 and 7 have more distinct vectors on the shared latent representation. By checking them, we can find the requirement 3, which asks for skills in ``Excel'' and ``Work'', is obviously about the basic professional tools required for this job. Meanwhile, the requirement 4 and 7, containing ``data analysis'', ``logic thinking'', ``business insight'', and ``judgement'', refer to the professional qualities for this job. In other words, these requirement items with distinct representations are highly relevant to the professional skills of this job. On the other hand, the representations of the requirement 1 and the requirement 5 are very vague. The first one refers to educational background and the fifth one is about working attitude. Obviously, in a high tech company, both of them are very common requirements. Thus we think their vague representations are also reasonable. Besides, we think the reason behind the vague representation of the requirement 2 is that compared with the terminology ``risk management'', those skill-related words can appear in job posting more frequently and thus can been learned more distinctly. Actually, only about 60 out of 15,039 jobs contain ``risk management'' in our dataset.

\textbf{Job Posting \#2.} Considering the keywords ``PHP'', ``lighttpd'', and ``MySQL'', this posting is for recruiting web development programmers. Actually, the requirement 4 therein, which is about the basic coding skills, has very distinct representations in this posting. It directly reflects the job content and the salience is reasonable. However, another technology-related requirement 2, which contains ``PHP'', ``lighttpd'', and ``NGINX'', are relatively vague. We think the reason behind this phenomenon is these web framework and specific skills are always being updated with rapid speed, thus solid foundation of coding is more important. In fact, by checking a lot of requirements, we find that the web-related requirement items~(e.g. for DJANGO and HTTP) often have more vague representations than those about basic coding skills, such as ``C/C++'', ``Python'', and ``Algorithm''.

\textbf{Candidate Resume \#1.} It is obvious that all experiences of this candidate are about risk management in banks. For example, the experience 2 is about developing ``a credit rating model to evaluate debtors'' and the experience 3 is to ``optimize the market risk policy system''. Obviously this candidate has rich experiences in risk management and it is reasonable that her representation learned by PJFNN is close to that of the job posting \#1. Actually, cosine similarities between this resume and the above two job postings are 0.89 and 0.12, respectively. This result further validates the effectiveness of our model.

\textbf{Candidate Resume \#2.} The overall representation of this resume is relatively vague. Meanwhile the representation of the experience 2 is distinct and is similar to the job postings \#2, which is about web development. Considering its top frequency words, containing ``Apache'', ``PHP'', and ``online'', are highly related to web technology, this similarity is reasonable. However, the experience 1 is so vague that its latent vector is close to neither of above job postings. Actually, the person described the project a lot but did not figure out her duty clearly. And PJFNN cannot extract much useful information. Thus, the similarities of this resume and both of the two jobs are not high~(e.g. 0.42 for job posting \#1 and 0.45 for job posting \#2).

\begin{figure*}[t]
\centering
\includegraphics[width=\textwidth]{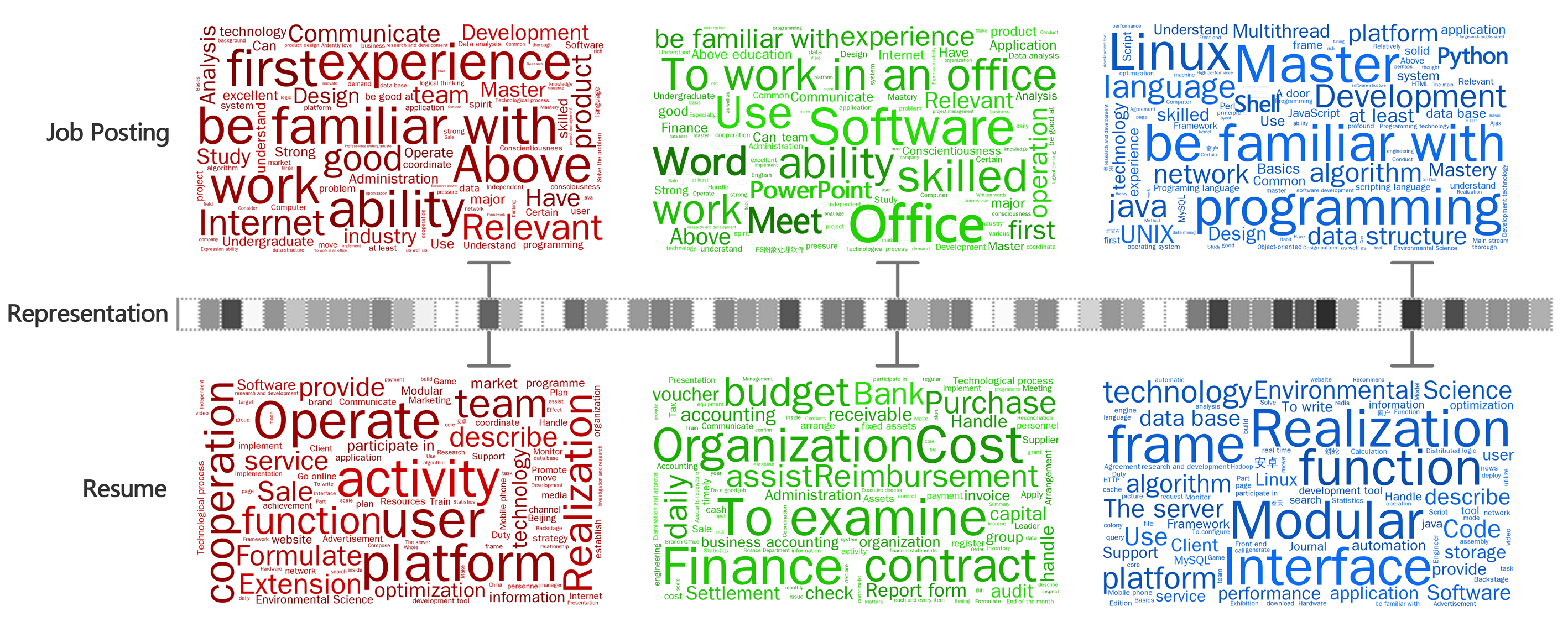}\\
\caption{The word clouds of three latent dimensions of representation learned by PJFNN, where the size of each keyword is proportional to its probabilities.}\label{vecKeywords}
\end{figure*}

For further demonstrating the interpretability of PJFNN, we show some keywords for three randomly selected dimensions of the latent representation learned in Figure~\ref{vecKeywords}. Due to the structure of PJFNN, we cannot directly match keywords into latent dimensions. Thus, we first selected those resumes (job postings) whose values in the given dimensions are very high and extracted the high-frequency words from them as the keywords of this dimension. It is obvious that the first dimension is related to program management, where job postings often contain keywords ``communicate'', ``design'', and ``product''. The frequencies of program management related words, such as ``operate'', ``cooperation'', and ``team'', are high in corresponding resumes. Meanwhile, the second dimension is more about administrative work, where job postings contain a lot of requirements for office software, such as ``Office'' and ``Word''. And the corresponding resumes, naturally containing ``finance'', ``assist'', and ``organization'', confirm this assumption. As for the third dimension, the high frequencies of coding related keywords, such as ``linux'', ``java'', and ``algorithm'', clearly validate its relationship to IT development. This relationship is also proven from the perspective of resumes, where ``modular'', ``interface'', and ``frame'' have very high probabilities.

\subsection{Empirical Studies}
In this subsection, we will further empirically study some real-world Person-Job Fit cases based on the results of PJFNN.

We randomly selected some job requirements from job postings and work experiences from resumes, and calculated their pairwise cosine similarities. The results are demonstrated in Table~\ref{tab:vectorItemCase}. We have highlighted the keywords in these work experience items manually. The requirement \#1, which contains ``C/C++ development'' and ``algorithms'', is obviously about coding skills and those work experience items with high similarities to it are also about coding. For example, the first work experience item has the highest similarity and it is about the file system development in Unix. The second one refers to ``thread scheduling, file system and memory management'' in Linux. Besides, although the third experience, which is about UI/UE, is related to software development, it cannot unveil the coding skills of this candidate directly. Thus we think its relatively low similarity is reasonable. Meanwhile, the other experiences, which are about education, data collection, or startups, have less relationship with coding and thus have low similarities.

The requirement \#2, which contains ``product development procedure'' and ``documenting'', is obviously for program management. Thus those work experience items about project management automatically have high similarities. In the first one, the candidate pointed out clearly she was responsible for the product in IBM and had rich experience in product management. The second one also figured out directly the corresponding candidate's duty as a program manager in RenRen mini version. Thus both of their similarities are high. Meanwhile the similarities of those work experience items about coding, such as the third and the forth one, are low accordingly. As for those IT-irrelevant experiences, such as the fifth and sixth one which are about education and resource management respectively, it is undoubted that their similarities are close to zero.

However, not all of job requirements can be modeled well in PJFNN. For example, the requirement \#3 is an education requirement, which widely appears in the job posting data. In the results, none of work experience items has high similarity to it. As mentioned before, we think this phenomenon should be attributed to almost every job in our dataset requires ``With bachelor degree or above''. Thus PJFNN cannot learn a distinct representation for this requirement. In a word, we think although PJFNN cannot learn good representations for all of requirements, the latent vectors of most resumes and job postings learned by PJFNN are meaningful generally, and can help to improve the effectiveness and efficiency of Person-Job Fit.

\section{Conclusion}
In this paper, we proposed a novel end-to-end model based on CNN, namely Person-Job Fit Neural Network (PJFNN), for matching a talent qualification to the requirement of a job. To be specific, PJFNN is a bipartite neural network which can effectively learn the joint representation of Person-Job fitness from historical job applications, thus it can project both job postings and candidates' resumes onto a shared latent representation. In particular, with the design of hierarchical representation structure, PJFNN can not only estimate whether a candidate fits a job, but also answer which requirements in the job posting are met by the candidate, through the measurement of distances between corresponding latent representations. Finally, we evaluated our model based on a large-scale real-world dataset collected from a high tech company in China. The extensive experiments clearly validate the performance of our model in terms of Person-Job Fit prediction, and demonstrate some interesting discoveries by visualizing the results obtained by PJFNN.

\begin{table*}[]
\scriptsize
\caption{Some matching results between job requirements and work experiences, where the skill related statements are highlighted for facilitating understanding.}
\begin{tabular}{|p{0.2\textwidth}|c|p{0.6\textwidth}|}
\hline
Concrete Job Requirement Items & Similarities & Concrete Work Experience Items \\
\hline
\multirow{6}{0.2\textwidth}{Requirement \#1: Proficient in C/C++ development, familiar with relevant algorithms} & 0.99779 & USD(Unified Storage Division) Senior Software Engineer. In charge of \textbf{file system development} of EMC middle-scale storage system. Have experience in many iterations. \textbf{Familiar with Unix file system, Common Block File System and 64-bit file system development.} Proficient in Volume....\\
\cline{2-3} & 0.99689 & Conducting Nachos OS kernel maintaince \textbf{on Linux with C++}. Primarily \textbf{be responsible for core modules like thread scheduling, file system and memory management}. Refined the algorithm of thread scheduling....\\
\cline{2-3} & 0.40335 & UI designer. Description: 1. Conducting \textbf{Graphic User Interface design} based on demands, for both mobile and TV client. 2. Established standards \& manners for UI design. 3. Provided impression drawings....\\
\cline{2-3} & 0.00056 & Youjian Jiaoshi, \textbf{Intern
Painting teacher} for kids between 4-10 years old. Giving lectures twice a week.\\
\cline{2-3} & 0.00053 & Supervisor of Shanghai Data Dept. Taking charge of \textbf{routine work}. Established the first off-site data collecting team. Be responsible for \textbf{data collecting outsourcing management}. Evaluating and testing data from multi-source.\\
\cline{2-3} & 0.00053 & Merchandising Manager. \textbf{Set up a company} with 2 partners in 2005, involving in the innovation and sell of measuring instrumentation chips. Gross income of 2006 was 2.8 million RMB, which climbed all the way to 25 million....\\
\hline
\multirow{6}{0.2\textwidth}{Requirement \#2: Able to schedule and design products, familiar with product development procedure and documenting.} & 0.86092 & IBM, Senior Software Engineer. Worked in IBM CDL for more than 5 years. Be \textbf{responsible for WebSphere product} line and get PMP certification. \textbf{Abundant experience in product development, test and management}....\\
\cline{2-3} & 0.82261 & Be \textbf{responsible for RenRen mini-version} from the very beginning, re-directing to the ``Social communication tools'' market. Rebuilt the 20\% popular functions(80\% usage covered) of RenRen client following....\\
\cline{2-3} & 0.31561 & Lego is an editor-oriented system, consisting video producing, topic customizing, page publishing and contract managing functions. This system includes \textbf{modules like jquery, jqueryui} and etc.\\
\cline{2-3} & 0.31574 & Joint with product lines of company, assisting to \textbf{solve problems with regard to user mode and kernel mode}. Make sure the innovated Linux-based products is commercially wide-used.\\
\cline{2-3} & 0.07274 & To \textbf{promote the scientific research and teaching combine}, and to culture the research abilities and team spirit of students, the school offers an innovative educational fieldwork project, relying on our research teams and....\\
\cline{2-3} & 0.07267 & TMS is a comprehensive intelligent traffic system. It integrates isometry devices and systems, in order to \textbf{resolve the isomerism of protocols and data}. I was \textbf{responsible for the resource management module}.\\
\hline
\multirow{6}{0.2\textwidth}{Requirement \#3: With bachelor degree or above} & 0.57259 & \textbf{Website Editor}. 1. Schedule the website pages structure, and plan for the contents. 2. Be responsible for mobile reading products. 3. Optimize product structure and funtions with cross-dept communications\\
\cline{2-3} & 0.26635 & In charge of chess\&card game opeartion and PC/mobile-client game-related translation for Facebook Japan. \textbf{Conducting data analysis} to enhance the DAU by 20\%, and the quarter gross income by 100\%.\\
\cline{2-3} & 0.17677 & This project is to break into the after-sale automobile service market, including preservation, repairment, rescue and second-hand sell, invested by CADIA. Be \textbf{responsible for the market investigation}.\\
\cline{2-3} & 0.14005 & Use \textbf{CNN to identify} the pattern of live eelworm. And help to enhance the robustness of real-time race.\\
\cline{2-3} & 0.05386 & USD(Unified Storage Division) Senior Software Engineer. In charge of \textbf{file system development} of EMC middle-scale storage system. Have experience in many iterations. \textbf{Familiar with Unix file system, Common Block File System and 64-bit file system development.} Proficient in Volume....\\
\cline{2-3} & 0.52233 & Be \textbf{responsible for RenRen mini-version} from the very beginning, re-directing to the ``Social communication tools'' market. Rebuilt the 20\% popular functions(80\% usage covered) of RenRen client following....
\\
\hline
\end{tabular}\label{tab:vectorItemCase}
\vspace{-3mm}
\end{table*}

\bibliographystyle{ACM-Reference-Format}

\begin{thebibliography}{34}


\ifx \showCODEN    \undefined \def \showCODEN     #1{\unskip}     \fi
\ifx \showDOI      \undefined \def \showDOI       #1{#1}\fi
\ifx \showISBNx    \undefined \def \showISBNx     #1{\unskip}     \fi
\ifx \showISBNxiii \undefined \def \showISBNxiii  #1{\unskip}     \fi
\ifx \showISSN     \undefined \def \showISSN      #1{\unskip}     \fi
\ifx \showLCCN     \undefined \def \showLCCN      #1{\unskip}     \fi
\ifx \shownote     \undefined \def \shownote      #1{#1}          \fi
\ifx \showarticletitle \undefined \def \showarticletitle #1{#1}   \fi
\ifx \showURL      \undefined \def \showURL       {\relax}        \fi
\providecommand\bibfield[2]{#2}
\providecommand\bibinfo[2]{#2}
\providecommand\natexlab[1]{#1}
\providecommand\showeprint[2][]{arXiv:#2}

\bibitem[\protect\citeauthoryear{Cheng, Xie, Chen, Agrawal, Choudhary, and
  Guo}{Cheng et~al\mbox{.}}{2013}]%
        {cheng2013jobminer}
\bibfield{author}{\bibinfo{person}{Yu Cheng}, \bibinfo{person}{Yusheng Xie},
  \bibinfo{person}{Zhengzhang Chen}, \bibinfo{person}{Ankit Agrawal},
  \bibinfo{person}{Alok Choudhary}, {and} \bibinfo{person}{Songtao Guo}.}
  \bibinfo{year}{2013}\natexlab{}.
\newblock \showarticletitle{Jobminer: A real-time system for mining job-related
  patterns from social media}. In \bibinfo{booktitle}{\emph{Proceedings of the
  19th ACM SIGKDD international conference on Knowledge discovery and data
  mining}}. ACM, \bibinfo{pages}{1450--1453}.
\newblock


\bibitem[\protect\citeauthoryear{Cho, van Merri{\"e}nboer, Bahdanau, and
  Bengio}{Cho et~al\mbox{.}}{2014}]%
        {cho2014properties}
\bibfield{author}{\bibinfo{person}{Kyunghyun Cho}, \bibinfo{person}{Bart van
  Merri{\"e}nboer}, \bibinfo{person}{Dzmitry Bahdanau}, {and}
  \bibinfo{person}{Yoshua Bengio}.} \bibinfo{year}{2014}\natexlab{}.
\newblock \showarticletitle{On the Properties of Neural Machine Translation:
  Encoder--Decoder Approaches}.
\newblock \bibinfo{journal}{\emph{Syntax, Semantics and Structure in
  Statistical Translation}} (\bibinfo{year}{2014}), \bibinfo{pages}{103}.
\newblock


\bibitem[\protect\citeauthoryear{Devlin, Zbib, Huang, Lamar, Schwartz, and
  Makhoul}{Devlin et~al\mbox{.}}{2014}]%
        {devlin2014fast}
\bibfield{author}{\bibinfo{person}{Jacob Devlin}, \bibinfo{person}{Rabih Zbib},
  \bibinfo{person}{Zhongqiang Huang}, \bibinfo{person}{Thomas Lamar},
  \bibinfo{person}{Richard Schwartz}, {and} \bibinfo{person}{John Makhoul}.}
  \bibinfo{year}{2014}\natexlab{}.
\newblock \showarticletitle{Fast and Robust Neural Network Joint Models for
  Statistical Machine Translation}. In \bibinfo{booktitle}{\emph{Proceedings of
  the 52th Annual Meeting of the Association for Computational Linguistics}}.
  \bibinfo{pages}{1370--1380}.
\newblock


\bibitem[\protect\citeauthoryear{Diaby, Viennet, and Launay}{Diaby
  et~al\mbox{.}}{2013}]%
        {diaby2013toward}
\bibfield{author}{\bibinfo{person}{Mamadou Diaby}, \bibinfo{person}{Emmanuel
  Viennet}, {and} \bibinfo{person}{Tristan Launay}.}
  \bibinfo{year}{2013}\natexlab{}.
\newblock \showarticletitle{Toward the next generation of recruitment tools: an
  online social network-based job recommender system}. In
  \bibinfo{booktitle}{\emph{Proceedings of the 2013 IEEE/ACM International
  Conference on Advances in Social Networks Analysis and Mining}}. ACM,
  \bibinfo{pages}{821--828}.
\newblock


\bibitem[\protect\citeauthoryear{Factbook}{Factbook}{2016}]%
        {bersin}
\bibfield{author}{\bibinfo{person}{Talent~Acquisition Factbook}.}
  \bibinfo{year}{2016}\natexlab{}.
\newblock
  \bibinfo{publisher}{http://blog.bersin.com/benchmarking-talent-acquisition-increasing-spend-cost-per-hire-and-time-to-fill/.
  Retrieved on 19 December 2017}.
\newblock


\bibitem[\protect\citeauthoryear{Hermann and Blunsom}{Hermann and
  Blunsom}{2014}]%
        {hermann2014multilingual2}
\bibfield{author}{\bibinfo{person}{Karl~Moritz Hermann} {and}
  \bibinfo{person}{Phil Blunsom}.} \bibinfo{year}{2014}\natexlab{}.
\newblock \showarticletitle{Multilingual Distributed Representations without
  Word Alignment}. In \bibinfo{booktitle}{\emph{Proceedings of ICLR}}.
\newblock


\bibitem[\protect\citeauthoryear{Holland}{Holland}{1973}]%
        {holland1973making}
\bibfield{author}{\bibinfo{person}{John~L Holland}.}
  \bibinfo{year}{1973}\natexlab{}.
\newblock \bibinfo{booktitle}{\emph{Making vocational choices: A theory of
  careers}}.
\newblock \bibinfo{publisher}{Prentice Hall}, \bibinfo{address}{Upper Saddle
  River, NJ}.
\newblock


\bibitem[\protect\citeauthoryear{Hong, Zheng, and Wang}{Hong
  et~al\mbox{.}}{2013}]%
        {hong2013dynamic}
\bibfield{author}{\bibinfo{person}{Wenxing Hong}, \bibinfo{person}{Siting
  Zheng}, {and} \bibinfo{person}{Huan Wang}.} \bibinfo{year}{2013}\natexlab{}.
\newblock \showarticletitle{Dynamic user profile-based job recommender system}.
  In \bibinfo{booktitle}{\emph{Computer Science \& Education (ICCSE), 2013 8th
  International Conference on}}. IEEE, \bibinfo{pages}{1499--1503}.
\newblock


\bibitem[\protect\citeauthoryear{Ioffe and Szegedy}{Ioffe and Szegedy}{2015}]%
        {ioffe2015batch}
\bibfield{author}{\bibinfo{person}{Sergey Ioffe} {and}
  \bibinfo{person}{Christian Szegedy}.} \bibinfo{year}{2015}\natexlab{}.
\newblock \showarticletitle{Batch normalization: Accelerating deep network
  training by reducing internal covariate shift}. In
  \bibinfo{booktitle}{\emph{International Conference on Machine Learning}}.
  \bibinfo{pages}{448--456}.
\newblock


\bibitem[\protect\citeauthoryear{Kalchbrenner, Grefenstette, Blunsom,
  Kartsaklis, Kalchbrenner, Sadrzadeh, Kalchbrenner, Blunsom, Kalchbrenner, and
  Blunsom}{Kalchbrenner et~al\mbox{.}}{2014}]%
        {kalchbrenner2014convolutional}
\bibfield{author}{\bibinfo{person}{Nal Kalchbrenner}, \bibinfo{person}{Edward
  Grefenstette}, \bibinfo{person}{Phil Blunsom}, \bibinfo{person}{Dimitri
  Kartsaklis}, \bibinfo{person}{Nal Kalchbrenner}, \bibinfo{person}{Mehrnoosh
  Sadrzadeh}, \bibinfo{person}{Nal Kalchbrenner}, \bibinfo{person}{Phil
  Blunsom}, \bibinfo{person}{Nal Kalchbrenner}, {and} \bibinfo{person}{Phil
  Blunsom}.} \bibinfo{year}{2014}\natexlab{}.
\newblock \showarticletitle{A Convolutional Neural Network for Modelling
  Sentences}. In \bibinfo{booktitle}{\emph{Proceedings of the 52nd Annual
  Meeting of the Association for Computational Linguistics}}. Association for
  Computational Linguistics, \bibinfo{pages}{212--217}.
\newblock


\bibitem[\protect\citeauthoryear{Kim}{Kim}{2014}]%
        {kim2014convolutional}
\bibfield{author}{\bibinfo{person}{Yoon Kim}.} \bibinfo{year}{2014}\natexlab{}.
\newblock \showarticletitle{Convolutional neural networks for sentence
  classification}. In \bibinfo{booktitle}{\emph{Proceedings of the 2014
  Conference on Empirical Methods in Natural Language Processing}}. Association
  for Computational Linguistics, \bibinfo{pages}{1746--1751}.
\newblock


\bibitem[\protect\citeauthoryear{Kingma and Ba}{Kingma and Ba}{2014}]%
        {Kingma2014Adam}
\bibfield{author}{\bibinfo{person}{Diederik Kingma} {and}
  \bibinfo{person}{Jimmy Ba}.} \bibinfo{year}{2014}\natexlab{}.
\newblock \showarticletitle{Adam: A method for stochastic optimization}.
\newblock \bibinfo{journal}{\emph{arXiv preprint arXiv:1412.6980}}
  (\bibinfo{year}{2014}).
\newblock


\bibitem[\protect\citeauthoryear{Lauly, Boulanger, and Larochelle}{Lauly
  et~al\mbox{.}}{2014}]%
        {Lauly2014Learning}
\bibfield{author}{\bibinfo{person}{Stanislas Lauly}, \bibinfo{person}{Alex
  Boulanger}, {and} \bibinfo{person}{Hugo Larochelle}.}
  \bibinfo{year}{2014}\natexlab{}.
\newblock \showarticletitle{Learning multilingual word representations using a
  bag-of-words autoencoder}.
\newblock \bibinfo{journal}{\emph{arXiv preprint arXiv:1401.1803}}
  (\bibinfo{year}{2014}).
\newblock


\bibitem[\protect\citeauthoryear{Li, Arya, Ha-Thuc, and Sinha}{Li
  et~al\mbox{.}}{2016}]%
        {Li2016How}
\bibfield{author}{\bibinfo{person}{Jia Li}, \bibinfo{person}{Dhruv Arya},
  \bibinfo{person}{Viet Ha-Thuc}, {and} \bibinfo{person}{Shakti Sinha}.}
  \bibinfo{year}{2016}\natexlab{}.
\newblock \showarticletitle{How to Get Them a Dream Job?: Entity-Aware Features
  for Personalized Job Search Ranking}. In
  \bibinfo{booktitle}{\emph{Proceedings of the 22th ACM SIGKDD international
  conference on Knowledge discovery and data mining}}.
  \bibinfo{pages}{501--510}.
\newblock


\bibitem[\protect\citeauthoryear{Lin, Zhu, Zuo, Zhu, Wu, and Xiong}{Lin
  et~al\mbox{.}}{2017}]%
        {Lin2017Collaborative}
\bibfield{author}{\bibinfo{person}{Hao Lin}, \bibinfo{person}{Hengshu Zhu},
  \bibinfo{person}{Yuan Zuo}, \bibinfo{person}{Chen Zhu},
  \bibinfo{person}{Junjie Wu}, {and} \bibinfo{person}{Hui Xiong}.}
  \bibinfo{year}{2017}\natexlab{}.
\newblock \showarticletitle{Collaborative Company Profiling: Insights from an
  Employee's Perspective}. In \bibinfo{booktitle}{\emph{Proceedings of AAAI}}.
  \bibinfo{pages}{1417--1423}.
\newblock


\bibitem[\protect\citeauthoryear{linkedIn Wikipedia}{linkedIn
  Wikipedia}{2017}]%
        {linkedinWiki}
\bibfield{author}{\bibinfo{person}{linkedIn Wikipedia}.}
  \bibinfo{year}{2017}\natexlab{}.
\newblock \bibinfo{publisher}{https://en.wikipedia.org/wiki/LinkedIn. Retrieved
  on 19 December 2017}.
\newblock


\bibitem[\protect\citeauthoryear{Lu, El~Helou, and Gillet}{Lu
  et~al\mbox{.}}{2013}]%
        {lu2013recommender}
\bibfield{author}{\bibinfo{person}{Yao Lu}, \bibinfo{person}{Sandy El~Helou},
  {and} \bibinfo{person}{Denis Gillet}.} \bibinfo{year}{2013}\natexlab{}.
\newblock \showarticletitle{A recommender system for job seeking and recruiting
  website}. In \bibinfo{booktitle}{\emph{Proceedings of the 22nd International
  Conference on World Wide Web}}. ACM, \bibinfo{pages}{963--966}.
\newblock


\bibitem[\protect\citeauthoryear{Malinowski, Keim, Wendt, and
  Weitzel}{Malinowski et~al\mbox{.}}{2006}]%
        {malinowski2006matching}
\bibfield{author}{\bibinfo{person}{Jochen Malinowski}, \bibinfo{person}{Tobias
  Keim}, \bibinfo{person}{Oliver Wendt}, {and} \bibinfo{person}{Tim Weitzel}.}
  \bibinfo{year}{2006}\natexlab{}.
\newblock \showarticletitle{Matching people and jobs: A bilateral
  recommendation approach}. In \bibinfo{booktitle}{\emph{Proceedings of the
  39th Annual Hawaii International Conference on System Sciences}}. IEEE.
\newblock


\bibitem[\protect\citeauthoryear{Mikolov, Chen, Corrado, and Dean}{Mikolov
  et~al\mbox{.}}{2013}]%
        {Mikolov2013Efficient}
\bibfield{author}{\bibinfo{person}{Tomas Mikolov}, \bibinfo{person}{Kai Chen},
  \bibinfo{person}{Greg Corrado}, {and} \bibinfo{person}{Jeffrey Dean}.}
  \bibinfo{year}{2013}\natexlab{}.
\newblock \showarticletitle{Efficient estimation of word representations in
  vector space}.
\newblock \bibinfo{journal}{\emph{arXiv preprint arXiv:1301.3781}}
  (\bibinfo{year}{2013}).
\newblock


\bibitem[\protect\citeauthoryear{Mimno, Wallach, Naradowsky, Smith, and
  McCallum}{Mimno et~al\mbox{.}}{2009}]%
        {mimno2009polylingual}
\bibfield{author}{\bibinfo{person}{David Mimno}, \bibinfo{person}{Hanna~M
  Wallach}, \bibinfo{person}{Jason Naradowsky}, \bibinfo{person}{David~A
  Smith}, {and} \bibinfo{person}{Andrew McCallum}.}
  \bibinfo{year}{2009}\natexlab{}.
\newblock \showarticletitle{Polylingual topic models}. In
  \bibinfo{booktitle}{\emph{Proceedings of the 2009 Conference on Empirical
  Methods in Natural Language Processing: Volume 2-Volume 2}}. Association for
  Computational Linguistics, \bibinfo{pages}{880--889}.
\newblock


\bibitem[\protect\citeauthoryear{Nair and Hinton}{Nair and Hinton}{2010}]%
        {nair2010rectified}
\bibfield{author}{\bibinfo{person}{Vinod Nair} {and}
  \bibinfo{person}{Geoffrey~E Hinton}.} \bibinfo{year}{2010}\natexlab{}.
\newblock \showarticletitle{Rectified linear units improve restricted boltzmann
  machines}. In \bibinfo{booktitle}{\emph{Proceedings of the 27th international
  conference on machine learning (ICML-10)}}. \bibinfo{pages}{807--814}.
\newblock


\bibitem[\protect\citeauthoryear{Paparrizos, Cambazoglu, and Gionis}{Paparrizos
  et~al\mbox{.}}{2011}]%
        {paparrizos2011machine}
\bibfield{author}{\bibinfo{person}{Ioannis Paparrizos},
  \bibinfo{person}{B~Barla Cambazoglu}, {and} \bibinfo{person}{Aristides
  Gionis}.} \bibinfo{year}{2011}\natexlab{}.
\newblock \showarticletitle{Machine learned job recommendation}. In
  \bibinfo{booktitle}{\emph{Proceedings of the fifth ACM Conference on
  Recommender Systems}}. ACM, \bibinfo{pages}{325--328}.
\newblock


\bibitem[\protect\citeauthoryear{Qin, Zhu, Xu, Zhu, Jiang, Chen, and Xiong}{Qin
  et~al\mbox{.}}{2018}]%
        {qin2018Enhancing}
\bibfield{author}{\bibinfo{person}{Chuan Qin}, \bibinfo{person}{Hengshu Zhu},
  \bibinfo{person}{Tong Xu}, \bibinfo{person}{Chen Zhu}, \bibinfo{person}{Liang
  Jiang}, \bibinfo{person}{Enhong Chen}, {and} \bibinfo{person}{Hui Xiong}.}
  \bibinfo{year}{2018}\natexlab{}.
\newblock \showarticletitle{Enhancing Person-Job Fit for Talent Recruitment: An
  Ability-aware Neural Network Approach}.
\newblock  (\bibinfo{year}{2018}).
\newblock


\bibitem[\protect\citeauthoryear{Robbins}{Robbins}{2001}]%
        {robbins2001organizational}
\bibfield{author}{\bibinfo{person}{Stephen~P Robbins}.}
  \bibinfo{year}{2001}\natexlab{}.
\newblock \bibinfo{booktitle}{\emph{Organizational behavior, 14/E}}.
\newblock \bibinfo{publisher}{Pearson Education India},
  \bibinfo{address}{Delhi}.
\newblock


\bibitem[\protect\citeauthoryear{Shen, Zhu, Zhu, Xu, Ma, and Xiong}{Shen
  et~al\mbox{.}}{2018}]%
        {shen2018joint}
\bibfield{author}{\bibinfo{person}{Dazhong Shen}, \bibinfo{person}{Hengshu
  Zhu}, \bibinfo{person}{Chen Zhu}, \bibinfo{person}{Tong Xu},
  \bibinfo{person}{Chao Ma}, {and} \bibinfo{person}{Hui Xiong}.}
  \bibinfo{year}{2018}\natexlab{}.
\newblock \showarticletitle{A Joint Learning Approach to Intelligent Job
  Interview Assessment}.
\newblock  (\bibinfo{year}{2018}).
\newblock


\bibitem[\protect\citeauthoryear{Sutskever, Vinyals, and Le}{Sutskever
  et~al\mbox{.}}{2014}]%
        {sutskever2014sequence}
\bibfield{author}{\bibinfo{person}{Ilya Sutskever}, \bibinfo{person}{Oriol
  Vinyals}, {and} \bibinfo{person}{Quoc~V Le}.}
  \bibinfo{year}{2014}\natexlab{}.
\newblock \showarticletitle{Sequence to sequence learning with neural
  networks}. In \bibinfo{booktitle}{\emph{Advances in neural information
  processing systems 27 (NIPS 2014)}}. \bibinfo{pages}{3104--3112}.
\newblock


\bibitem[\protect\citeauthoryear{Vinyals, Kaiser, Koo, Petrov, Sutskever, and
  Hinton}{Vinyals et~al\mbox{.}}{2015}]%
        {vinyals2015grammar}
\bibfield{author}{\bibinfo{person}{Oriol Vinyals}, \bibinfo{person}{{\L}ukasz
  Kaiser}, \bibinfo{person}{Terry Koo}, \bibinfo{person}{Slav Petrov},
  \bibinfo{person}{Ilya Sutskever}, {and} \bibinfo{person}{Geoffrey Hinton}.}
  \bibinfo{year}{2015}\natexlab{}.
\newblock \showarticletitle{Grammar as a foreign language}. In
  \bibinfo{booktitle}{\emph{Advances in Neural Information Processing Systems
  28 (NIPS 2015)}}. \bibinfo{pages}{2773--2781}.
\newblock


\bibitem[\protect\citeauthoryear{Vu, Adel, Gupta, et~al\mbox{.}}{Vu
  et~al\mbox{.}}{2016}]%
        {vu2016combining}
\bibfield{author}{\bibinfo{person}{Ngoc~Thang Vu}, \bibinfo{person}{Heike
  Adel}, \bibinfo{person}{Pankaj Gupta}, {et~al\mbox{.}}}
  \bibinfo{year}{2016}\natexlab{}.
\newblock \showarticletitle{Combining Recurrent and Convolutional Neural
  Networks for Relation Classification}. In
  \bibinfo{booktitle}{\emph{Proceedings of NAACL-HLT}}.
  \bibinfo{pages}{534--539}.
\newblock


\bibitem[\protect\citeauthoryear{Xu, Yu, Yang, Xiong, and Zhu}{Xu
  et~al\mbox{.}}{2016}]%
        {xu2016talent}
\bibfield{author}{\bibinfo{person}{Huang Xu}, \bibinfo{person}{Zhiwen Yu},
  \bibinfo{person}{Jingyuan Yang}, \bibinfo{person}{Hui Xiong}, {and}
  \bibinfo{person}{Hengshu Zhu}.} \bibinfo{year}{2016}\natexlab{}.
\newblock \showarticletitle{Talent Circle Detection in Job Transition
  Networks}. In \bibinfo{booktitle}{\emph{Proceedings of the 22th ACM SIGKDD
  international conference on Knowledge discovery and data mining}}.
  \bibinfo{pages}{655--664}.
\newblock


\bibitem[\protect\citeauthoryear{Xu, Zhu, Zhu, Li, and Xiong}{Xu
  et~al\mbox{.}}{2018}]%
        {xu2018measuring}
\bibfield{author}{\bibinfo{person}{Tong Xu}, \bibinfo{person}{Hengshu Zhu},
  \bibinfo{person}{Chen Zhu}, \bibinfo{person}{Pan Li}, {and}
  \bibinfo{person}{Hui Xiong}.} \bibinfo{year}{2018}\natexlab{}.
\newblock \showarticletitle{Measuring the Popularity of Job Skills in
  Recruitment Market: A Multi-Criteria Approach}.
\newblock  (\bibinfo{year}{2018}).
\newblock


\bibitem[\protect\citeauthoryear{Yin, Kann, Yu, and Sch{\"u}tze}{Yin
  et~al\mbox{.}}{2017}]%
        {yin2017comparative}
\bibfield{author}{\bibinfo{person}{Wenpeng Yin}, \bibinfo{person}{Katharina
  Kann}, \bibinfo{person}{Mo Yu}, {and} \bibinfo{person}{Hinrich Sch{\"u}tze}.}
  \bibinfo{year}{2017}\natexlab{}.
\newblock \showarticletitle{Comparative Study of CNN and RNN for Natural
  Language Processing}.
\newblock \bibinfo{journal}{\emph{arXiv preprint arXiv:1702.01923}}
  (\bibinfo{year}{2017}).
\newblock


\bibitem[\protect\citeauthoryear{Zhang, Zhou, Ma, Chen, Zhang, and
  Agarwal}{Zhang et~al\mbox{.}}{2016}]%
        {zhang2016glmix}
\bibfield{author}{\bibinfo{person}{XianXing Zhang}, \bibinfo{person}{Yitong
  Zhou}, \bibinfo{person}{Yiming Ma}, \bibinfo{person}{Bee-Chung Chen},
  \bibinfo{person}{Liang Zhang}, {and} \bibinfo{person}{Deepak Agarwal}.}
  \bibinfo{year}{2016}\natexlab{}.
\newblock \showarticletitle{GLMix: Generalized Linear Mixed Models For
  Large-Scale Response Prediction}. In \bibinfo{booktitle}{\emph{Proceedings of
  the 22nd ACM SIGKDD International Conference on Knowledge Discovery and Data
  Mining}}. ACM, \bibinfo{pages}{363--372}.
\newblock


\bibitem[\protect\citeauthoryear{Zhang, Yang, and Niu}{Zhang
  et~al\mbox{.}}{2015}]%
        {zhang2014research}
\bibfield{author}{\bibinfo{person}{Yingya Zhang}, \bibinfo{person}{Cheng Yang},
  {and} \bibinfo{person}{Zhixiang Niu}.} \bibinfo{year}{2015}\natexlab{}.
\newblock \showarticletitle{A Research of Job Recommendation System Based on
  Collaborative Filtering}. In \bibinfo{booktitle}{\emph{Proceedings of the
  Seventh International Symposium on Computational Intelligence and Design}}.
  \bibinfo{pages}{533--538}.
\newblock


\bibitem[\protect\citeauthoryear{Zhu, Zhu, Xiong, Ding, and Xie}{Zhu
  et~al\mbox{.}}{2016}]%
        {zhu2016recruitment}
\bibfield{author}{\bibinfo{person}{Chen Zhu}, \bibinfo{person}{Hengshu Zhu},
  \bibinfo{person}{Hui Xiong}, \bibinfo{person}{Pengliang Ding}, {and}
  \bibinfo{person}{Fang Xie}.} \bibinfo{year}{2016}\natexlab{}.
\newblock \showarticletitle{Recruitment Market Trend Analysis with Sequential
  Latent Variable Models}. In \bibinfo{booktitle}{\emph{Proceedings of the 22th
  ACM SIGKDD international conference on Knowledge discovery and data mining}}.
  \bibinfo{pages}{383--392}.
\newblock


\end{thebibliography}


\end{document}